\documentclass[12pt]{article}
\usepackage{graphicx}
\newcommand{\Mgf}{\ensuremath{\mathcal{M}}}

\begin{document}
\title{A transfer matrix approach to the enumeration of plane meanders}
\author{ 
Iwan Jensen\thanks{e-mail: I.Jensen@ms.unimelb.edu.au}, \\
Department of Mathematics and Statistics, \\
The University of Melbourne, \\
 Victoria 3010, Australia}
\maketitle
\bibliographystyle{plain}
\begin{abstract}
A closed plane meander of order $n$ is a closed self-avoiding curve
intersecting an infinite line $2n$ times. Meanders are considered 
distinct up to any smooth deformation leaving the line fixed. We have 
developed an improved algorithm, based on transfer matrix methods, for the 
enumeration  of plane meanders. While the algorithm has exponential
complexity, its rate of growth is much smaller than that of
previous algorithms. The algorithm is easily modified to enumerate 
various systems of closed meanders, semi-meanders, open meanders and
many other geometries.
\end{abstract}

\section{Introduction}

Meanders \cite{LZ} form a set of combinatorial problems concerned with the 
enumeration of self-avoiding curves crossing a line through a given number 
of points. Meanders are considered distinct up to any smooth deformation. 
This problem seems to date back at least to work of 
Poincar\'e on differential geometry \cite{Poincare}. Since then it has from 
time to time been studied by mathematicians in various contexts such as the 
folding of a strip of stamps \cite{Touchard,Koehler} or folding of maps 
\cite{Lunnon}. More  recently it has been related to enumerations of ovals 
in planar algebraic curves \cite{Arnold} and the classification of 
3-manifolds \cite{KS}. During the last decade or so has it has received 
considerable attention in other areas of science. In computer science 
meanders are related to the sorting of Jordan sequences \cite{HMRT} and have 
been used for lower bound arguments \cite{AM}. In physics meanders are 
relevant to the study of compact foldings of polymers \cite{FGG1,FGG2}, 
properties of the Temperley-Lieb algebra \cite{FGG3,Francesco1}, 
matrix models \cite{Makeenko,SS,Francesco2}, and defects in liquid
crystals and $2+1$ dimensional gravity \cite{Kholodenko}.

The difficulty in the enumeration of most interesting combinatorial 
problems is that, computationally, they are of exponential complexity. 
That is to say, the time it takes to calculate the first $n$ 
terms in the generating function grows asymptotically as $\lambda^n$, 
where $\lambda > 1$ is the growth rate. Initial efforts at computer 
enumeration of meanders were based on direct 
counting. Independently, Koehler \cite{Koehler} and Lunnon \cite{Lunnon},
studied the number of ways of folding a strip of stamps (or a map) 
of length $n$, and published result up to, respectively, $n=16$
and 24. Lando and Zvonkin \cite{LZ} studied closed meanders, 
open meanders and multi-component systems of closed meanders,
and calculated the number of open meanders up to $n=26$ and
the number of closed meanders up to $n=14$. The calculation of the number 
of closed meanders was subsequently extended up to $n=16$ by 
Pratt \cite{Pratt}. Di Francesco {\em et al.} studied semi-meanders, a 
problem equivalent to the stamp folding problem \cite{FGG1}, and extended
the calculation to $n=29$ \cite{FGG2}, in addition they studied numerous
other problems including that of multi-component systems of
semi-meanders.

Few exact and mathematically rigorous results have been obtained for 
any of the many meander problems. However, in a recent paper it was 
conjectured that some of the meander problems can be related to a 
gravitational version of a certain loop model \cite{FGG4}.
From the conformal field theory of the model, conjectures were proposed 
for the exact critical exponent of closed and open meanders,
$\alpha= (29+\sqrt{145})/12 = 3.4201328\ldots$, 
as well as the exponent for semi-meanders,
$\overline{\alpha}= 1+\sqrt{11}(\sqrt{29}+\sqrt{5})/24 = 2.0531987\ldots$.
This work has recently been extended to multi-component systems of 
closed and semi-meanders \cite{FGJ2000} and to various other geometries. 
Conjectures were given for the critical exponents as functions of the 
loop-fugacity $q$. These were checked numerically \cite{FGJ2000} and found to 
be correct within numerical error. In a recent paper \cite{JG2000}
we analysed extended series for the meander generating functions.
Using the numerical technique of differential approximants \cite{Guttmann89}
we obtained accurate estimates for the exponents and found that the 
conjecture for the semi-meander exponent is unlikely to be correct, while 
the conjecture for closed meanders is just consistent with the results
from the analysis.

The purpose of this paper is to give a detailed description of the
new and improved algorithm used to derive the series studied in
\cite{JG2000}. While the algorithm still has
exponential complexity, the growth rate is much smaller
than that experienced with direct counting, and consequently the
calculation can be carried much further. The algorithm is easily modified to 
enumerate various multi-component systems of closed meanders, semi-meanders 
or open meanders. In particular we have extended the calculation
for closed meanders up to $n=24$, for open meanders up to $n=43$, and
for semi-meanders up to $n=45$.

In section~\ref{sec:def} we shall briefly describe some meander problems
and define the meandric numbers. Section~\ref{sec:enum} contains a
detailed description of the transfer matrix algorithm for the enumeration 
of closed meanders and an outline of generalisations to other problems.
Finally we give our conclusions in section~\ref{sec:conclusion}.

\section{Definitions of meanders \label{sec:def}}

\begin{figure}
\begin{center}
\includegraphics{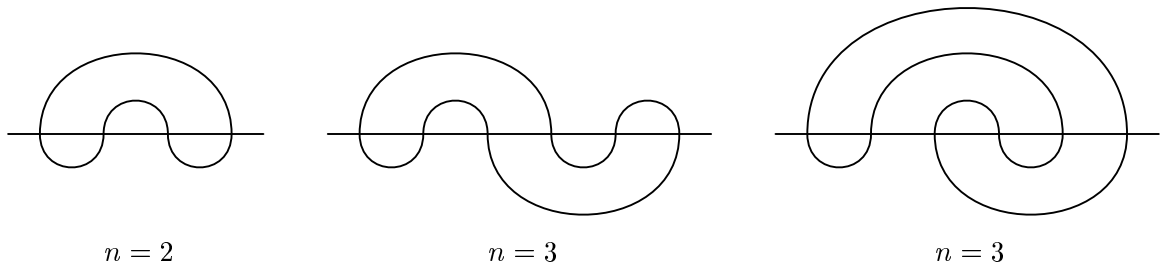}
\end{center}
\caption{\label{fig:meanclose} A few examples of closed meanders of
order 2 and 3, respectively. }
\end{figure}

A {\em closed meander} of order $n$ is a closed self-avoiding curve 
crossing an infinite line $2n$ times (see figure~\ref{fig:meanclose}). 
The meandric number $M_n$ is simply the number of such meanders distinct 
up to smooth transformations. Note that each meander forms a single 
connected curve. The number of closed meanders is expected to grow 
exponentially, with a sub-dominant term given by a critical exponent,
$M_n \sim  C R^{2n}/n^{\alpha}$. The exponential growth constant $R$
is often called the {\em connective constant}.
The generating function  is expected to behave as 

\begin{equation}\label{eq:meangen}
\Mgf(x) = \sum_{n=1}^{\infty} M_n x^n \sim A(x)(1 - R^2 x)^{\alpha-1},
\end{equation}
and hence have
a singularity at $x_c=1/R^2$ with exponent $\alpha -1$.

\begin{figure}[h]
\begin{center}
\includegraphics{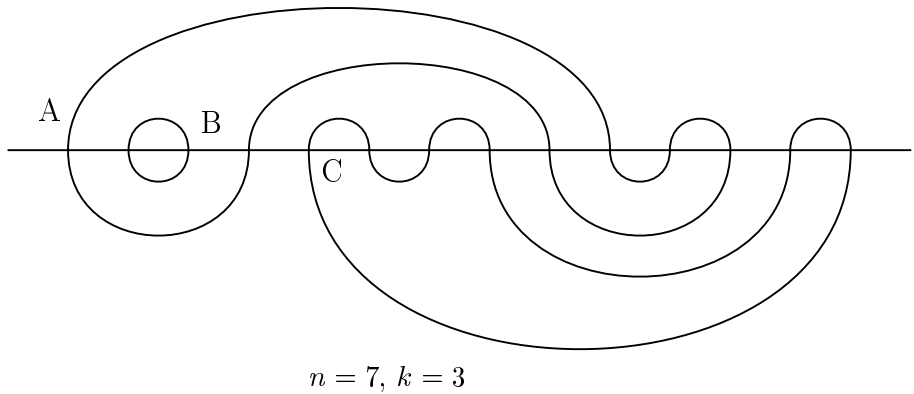}
\end{center}
\caption{\label{fig:meansys} An example of a system of closed meanders 
of order 7 with 3 components. }
\end{figure}

We can extend the definition to {\em multi-component systems of closed 
meanders}, where we allow configurations with several disconnected 
closed meanders. The meandric numbers $M_n^{(k)}$ are the number of 
meanders with $2n$ crossings and $k$ components, and we thus obtain 
the more general generating function:

\begin{equation}
\Mgf (x,q) = \sum_{n=1}^{\infty} \sum_{k=1}^n M_n^{(k)} x^nq^k.
\end{equation}
Obviously, $M_n=M_n^{(1)}$, and 
$\Mgf (x)=\partial \Mgf (x,q)/\partial q|_{q=0}$.
An example of a multi-component meander system with several pieces 
is shown in figure~\ref{fig:meansys}. In addition one could also look at 
further restrictions on these systems. Two obvious examples would be
to exclude configurations where meanders are nested within one another
or where meanders can be separated from  one another, i.e., all smaller
meanders are completely enclosed within a larger meander.

\begin{figure}[h]
\begin{center}
\includegraphics{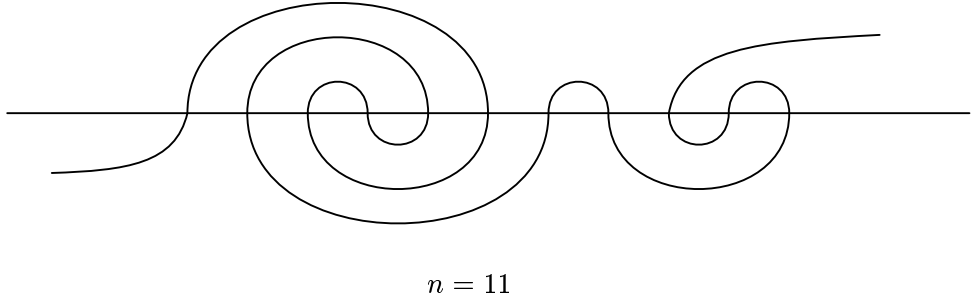}
\end{center}
\caption{\label{fig:meanopen} An example of an open meander of
order 11. }
\end{figure}

An {\em open meander} of order $n$ is a self-avoiding curve running from 
west to east while crossing an infinite line $n$ times (see
figure~\ref{fig:meanopen}). The number
of such curves is $m_n$ and we can define a generating function for
this problem in analogy with (\ref{eq:meangen}). It should be noted
\cite{LZ} that $M_n = m_{2n-1}$, and hence the critical exponent is 
identical to that of closed meanders and the connective constant is $R$.

\begin{figure}[h]
\begin{center}
\includegraphics{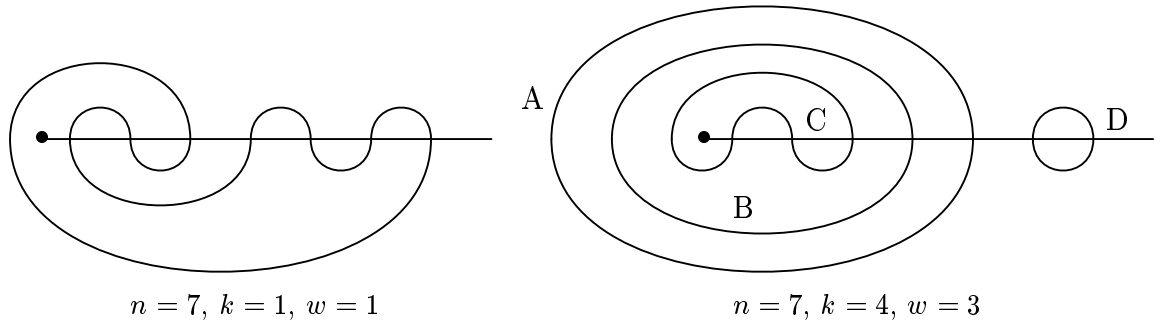}
\end{center}
\caption{\label{fig:meansemi} Two examples of semi-meanders.
The first of these has one component, crosses the line 7 times
and winds around the origin once, while the second has 4 components
crosses the line 7 times and winds around the origin 3 times.}
\end{figure}

Finally, instead of looking at crossings of an infinite line
one could consider a semi-infinite line and allow the curve to wind
around the end-point of the line \cite{FGG1}. A {\em semi-meander} of 
order $n$ is a closed self-avoiding curve crossing the semi-infinite 
line $n$ times. The number of semi-meanders of order $n$ is denoted by 
$\overline{M}_n\sim C' \overline{R}/n^{\overline{\alpha}}$ 
and we define a generating function as in (\ref{eq:meangen}).
In this case a further interesting generalisation is to study the number
of semi-meanders $\overline{M}_n(w)$ which wind around the  end-point of 
the line exactly $w$ times. Again we could also study systems of 
multi-component semi-meanders according to the number of independent 
meanders. Two semi-meanders are shown in figure~\ref{fig:meansemi}.

\section{Enumeration of meanders \label{sec:enum}}

\begin{figure}[h]
\begin{center}
\includegraphics[scale=0.8]{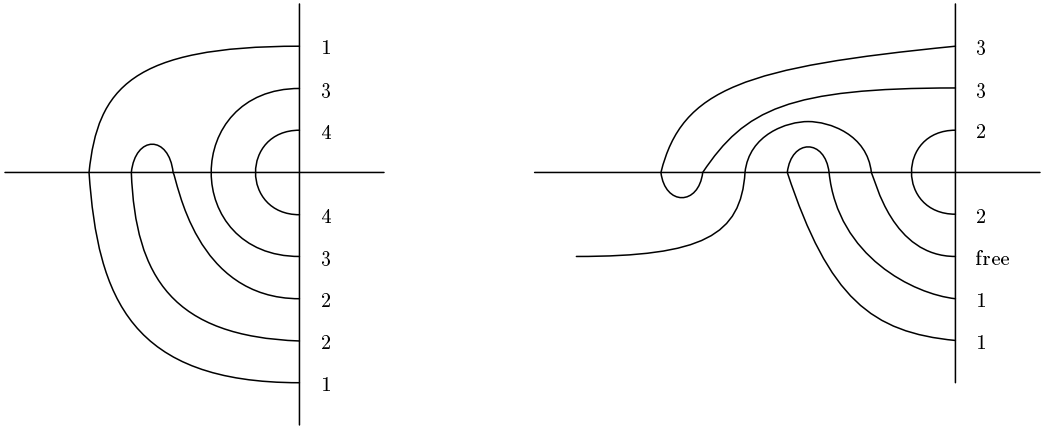}
\end{center}
\caption{\label{fig:intersect}
Examples of loop-configurations along the boundary (vertical line) 
during the transfer matrix calculation for closed meanders (left panel) 
and open meanders (right panel).  
Numbers along the boundary give a possible encoding of the
loop-ends in the partially completed meander.}
\end{figure}

The method used to enumerate meanders is based on the transfer matrix
algorithm devised by Enting \cite{Enting} in his pioneering work on the 
enumeration of self-avoiding polygons. Derrida \cite{Derrida} used a 
similar algorithm to study self-avoiding walks. The transfer matrix
technique involves drawing a boundary line perpendicular to the infinite 
line. The intersection between the boundary and a given meander 
results in a set of loop-ends. Each loop-end is connected 
(to the left of the boundary) to at most one other loop-end. In the case of 
closed meanders the matching is perfect and each loop-end is connected to 
exactly one other loop-end. In the case of open meanders there is in addition
one and only one loop-end which is free and therefore not connected to 
any other loop-ends, as illustrated in figure~\ref{fig:intersect}.
In addition to the information describing the configuration of loop-ends,
and how they are connected, we need to know where the infinite line is 
situated within the loop-ends. This can be done simply by specifying how 
many loop-ends lie beneath the infinite line. For each such configuration 
we keep count of all the possible (partially completed) meanders which gives 
rise to that particular configuration of loop-ends. Meanders can then be 
enumerated by successive moves of the boundary line, so that exactly one 
crossing is added with each move. An extra crossing is added {\em either}
by putting in a new loop across the infinite line {\em or} by taking
an existing loop-end immediately above/below the line and dragging
it to the other side. 

These general remarks hold for all the meander enumeration problems. In the 
following we give a detailed description of the algorithm used in the 
enumeration of closed connected meanders. Afterwards we describe how to 
generalise the method to other meander problems.

\subsection{Enumeration of closed connected meanders
\label{sec:enumclosed}}

\begin{figure}[h]
\begin{center}
\includegraphics[scale=0.8]{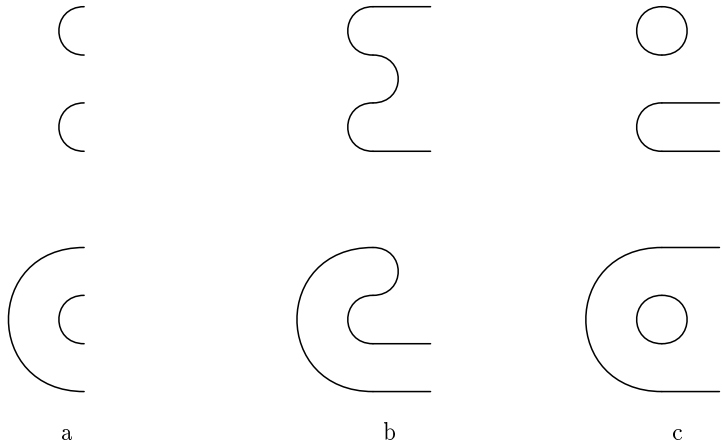}
\end{center}
\caption{\label{fig:looppair} Illustration of how two partial loops
can be placed (a), how they can be connected to from a single
loop (b), and connections leading to graphs with more than one
component (c). }
\end{figure}

In the enumeration of closed connected meanders the two major constraints 
which must be observed are self-avoidance and the constraint that all 
meanders constructed during the calculation must consist of a single
connected component. As we move the boundary line, the partially
completed meanders will consist of a number of disjoint loop segments,
which must be connected to each other if a valid closed meander is to
be produced. A pair of loops can be placed relative to one another in two 
distinct ways, namely, side by side or nested, as shown in the left panel 
of figure~\ref{fig:looppair}. In each case it is possible to connect the
loop-ends so as to form a single loop (middle panel) or so as to form
graphs with two separate components (right panel). It is connections
equivalent to these latter cases which we must avoid. So the constraint, 
which must be observed in order to avoid separate components, is that a
loop can be closed on itself only if the boundary intersects no other 
loops.

To avoid loops closing on themselves we have to label the loop-ends
so we can keep track of how they are connected to one another.
One scheme would be to assign a unique label to each loop as indicated
in figure~\ref{fig:intersect}. However, due to the two-dimensional nature 
of the problem and since the curve making up a meander is self-avoiding, 
there is a scheme better suited to a computer algorithm. Each configuration 
of loop-ends can be represented by an ordered set of states $\{\sigma_i\}$, 
where

\begin{equation}\label{eq:states}
\sigma_i  = \left\{ \begin{array}{rl}
0 &\;\;\; \mbox{lower end of a loop}, \\
1 &\;\;\; \mbox{upper end of a loop}. \end{array} \right.
\end{equation}
\noindent
It is easy to see that this encoding uniquely describes which
loop-ends are connected. In order to find the upper loop-end,
matching a given lower end, we start at the lower end and work upwards  
in the configuration counting the number of `0's and `1's we pass
(the `0' of the initial lower end is {\em not} included in the count).  
We stop when the number of `1's  exceeds the number of  `0's. This 
`1'  marks the matching upper end of the loop.
It is worth noting that there are some restrictions on the possible 
configurations. Firstly, every lower loop-end must have a corresponding 
upper end, and it is therefore clear that the total number of `0's  
is equal to the total number of `1's. Secondly, as we look through the 
configuration starting from the bottom the number of `0's is never 
smaller than the number of `1's. Those familiar with algebraic languages 
will immediately recognise that each configuration of labelled
loop-ends forms a Dyck word (see \cite{DV}).

\subsubsection{The transfer matrix algorithm}

The total configuration of loop-ends and their placement relative
to the infinite line can thus be described by a pair of integers $(h,S)$, 
where $h$ is the number of loop-ends below the infinite line and $S$ 
is the integer whose binary representation corresponds to the configuration 
of loop-ends. We shall call such a $(h,S)$-pair a {\em signature}, and in 
practise we represent it by a 64-bit integer with the first 6 bits coding 
$h$ and the remaining bits coding $S$. In the following we shall
often explicitly write out the binary representation, 
$\{ b_0 b_1 \ldots b_n \}$ of $S$, and use the notation $\{S_1 S_2 \}$
to mean a configuration of loop-ends obtained by concatenating the
binary strings $S_1$ and $S_2$. 

\begin{figure}[h]
\begin{center}
\includegraphics{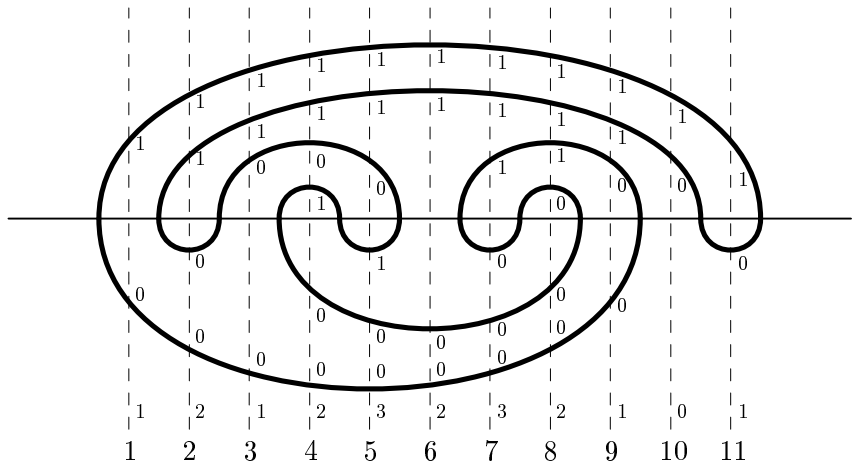}
\end{center}
\caption{\label{fig:transfer}
Positions of the boundaries (dashed lines) during the transfer matrix
calculation.  Numbers along the boundaries give the encoding of the 
loop structure of the intersection with the partially completed meander 
to the left of the boundaries.}
\end{figure}

The algorithm for enumerating closed connected meanders is:

\begin{enumerate}
\item
Set the maximum order $n$ of the meandric numbers we wish to calculate.
Start with the signature $(1,\{01\})$ with a count of 1, that is one loop 
crossing the infinite line. Set the number of crossings $n_c=1$. 
Set the meandric number $M_1 = 1$.

\item Next move the boundary one step ahead and add one more
crossing, $n_c=n_c+1$, to all existing signatures. 
A new crossing is added either by putting in an 
additional loop or by dragging an existing loop-end 
(immediately above or below) across the infinite line.

\begin{description}

\item[Adding:] A new loop is inserted simply by taking an existing 
{\em source} signature $(h,\{S_1 S_2 \})$, where $S_1$ ($S_2$) is the 
configuration  of loop-ends below  (above) the infinite line, and creating 
a new {\em target} signature $(h+1,\{S_1 01 S_2 \})$. The count of the 
source is added to the count of the target. This first type of move is 
illustrated in figure~\ref{fig:transfer} where in moving to position 2 
we generate the target signature $(2,\{0011\})$ from the source $(1,\{01\})$. 
Additional loops are also inserted while moving to positions 4 and 7.

\item[Dragging:] As we cross the infinite line with an existing loop-end we
{\em either} leave it at that {\em or} we may be allowed to connect the 
loop-end to the loop-end on the other side of the infinite line. 

\begin{description}

\item[No connection:]  From the source $(h,S)$ we produce the
the two targets $(h-1,S)$  and $(h+1,S)$, by dragging
a loop-end from below to above and vice versa.  
Both these updates are allowed provided $h-1 \geq 0$ 
and $h+1 \leq 2m_1$, where $m_1$ is the number of `1's in $S$.
For $h=0$ only the target $(1,S)$ is 
allowed and for $h=2m_1$ only the target $(h-1,S)$ is allowed.
Examples of these moves are given in figure~\ref{fig:transfer} 
when moving to positions 3, 5, 8 and 11.

\item[Connecting loop-ends:] There are four distinct cases depending
on whether the loop-ends below and above are of type `0' or
`1'.

\begin{description}

\item[Case 00:] In this case we connect a lower loop-end from below the 
line to a lower loop-end above the line. From a source $(h,\{S_100S_2\})$ we
generate the target  $(h-1,\{S_1\widehat{S_2}\})$, where the symbol
$\widehat{S_2}$ indicates that this string is changed via further processing. 
This is so because by connecting the two lower loop-ends an upper loop-end 
elsewhere in the old configuration $S_2$ becomes a lower loop-end in the 
new configuration $\widehat{S_2}$. An example of this type of relabelling 
is shown in figure~\ref{fig:transfer} where in the move to position 9 we see 
that the signature $(2,\{000111\})$ before the step becomes the configuration
$(1,\{0011\})$ after the step. That is, the upper end of the third loop
before the step becomes the lower end of the second loop after the step.  
In general the nesting of loops could be more complicated
and the general rule for the relabelling of the configuration is
as follows: When connecting two `0's  we work upwards  
in the configuration, counting the number of `0's and `1's we pass until 
the number of `1's  exceeds the number of  `0's. This 
`1'  is the matching end of the inner loop and it should now be changed
to a `0', thus becoming the lower end of the outer loop
(drawing a few further figures should make this relabelling clearer).

\item[Case 10:] In this case we connect an upper loop-end from below 
to a lower loop-end above. So from a source $(h,\{S_110S_2\})$ we
generate the target  $(h-1,\{S_1S_2\})$.

\item[Case 01:] This is never allowed since it would result in
a closed loop and thus generate graphs with separate components.
The only exception is when there are no other loop-ends in $S$, but
this case is dealt with in 3.

\item[Case 11:] In this case we connect an upper loop-end from below the line
to an upper loop-end above the line. From a source $(h,\{S_111S_2\})$ we
generate the target  $(h-1,\{\widehat{S_1}S_2\})$. 
The rule for the relabelling of $S_1$ $\widehat{S_1}$  is similar
to the case `00', but we work downwards in the string $S_1$ until we find 
the unmatched lower loop-end, which is then changed to an upper loop-end.

\end{description}

\end{description}

\end{description}

Note that all of the above moves may be allowed. So from a given
source we can generate up to four targets, by adding a new loop, by 
dragging a loop-end from below to above the infinite line, doing the 
reverse, or by connecting two loop-ends across the infinite
line. As we move along and generate new target signatures their counts 
are calculated by adding up the count for the various source signatures
which could generate that target. For example the target $(2,\{0011\})$ is 
generated from the sources $(1,\{01\})$, $(1,\{0011\})$, $(3,\{0011\})$, and
$(3,\{001011\})$, by, respectively, putting in an additional loop, moving a 
loop-end below the line, moving a loop-end above the line and connecting
two loop-ends across the line. 

\item If $n_c$ is odd then set $j=(n_c+1)/2$ and extract the meandric number
$M_j$ as the count of the signature $(1,\{01\})$. This is the only case
in which we are allowed to close a loop. Doing so obviously adds one more
crossing.

\item If $n_c<2N-1$ go to 2.

\end{enumerate}  

Not all possible signatures that can be generated in a calculation
to order $n$ are actually required. The main restriction is that no 
meanders should have more than $2n$ crossings. Since each move adds one
more crossing and reduces the number of loop-ends above/below the infinite 
line by at most one, it is clear that for a given signature we have to add 
at least $n_a=\max (h,2m_1-h)$ additional crossings in order to produce a 
closed meander. Thus if, for a given signature, $n_c+n_a>2n$, we can 
discard the signature since it would contribute only to a meandric number
exceeding the order to which we wish to carry out the calculation.
Further savings of a factor of almost 2 is  obtained by using the 
symmetry with respect to reflection in the infinite line. A further factor
of approximately 2 is obtained as follows. Note that $n_a$ is the minimum
number of additional crossings and that for some signatures further
crossings are needed. The most obvious case is when the loop-ends
above and below the infinite line are connected to one another. In
this case we cannot connect the two ends and first we have to
move one of the loop-ends across the line. So when $h=m_1$, this
results in at least two extra crossings. In the general case one can
readily write an algorithm to count the actual number of additional
crossings required, and as stated above this results in a saving
of close to 2 in the number of signatures one need retain.    

\subsection{Generalisations to other meander problems}

\subsubsection{Multi-component systems of closed meanders}

As we noted above  connecting a `0' below the line to a `1' above the line 
results in a closed loop. Failure to observe the restriction on this closure
would result in graphs with disconnected components, either one closed
meander over another or one closed meander within another. Obviously these 
are just the types of graphs required in order to enumerate multi-component 
systems of closed meanders. So by noting that each such closure adds one more
component it is straightforward to generalise the algorithm to enumerate 
systems of closed meanders. The only major change is that, rather than
just storing the number of partially completed meanders, for each signature 
we have to store a generating function, that is a polynomial giving 
the number of partially completed meanders with $k$ components, where
$1 \leq k \leq n$.  

\subsubsection{Open meanders}

\begin{figure}[h]
\begin{center}
\includegraphics[scale=0.8]{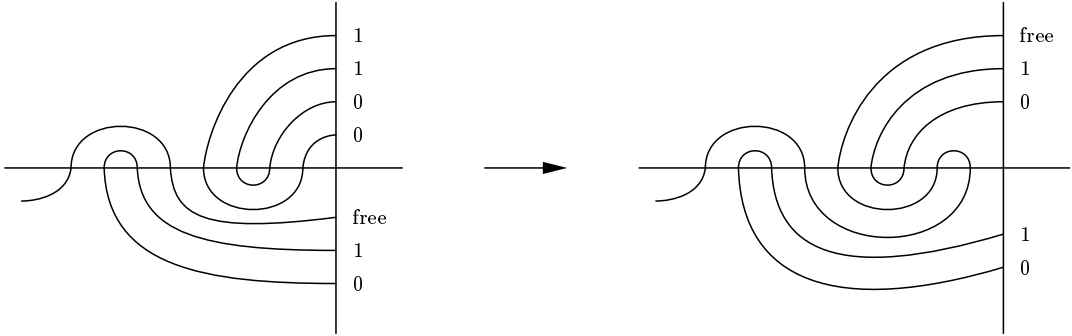}
\end{center}
\caption{\label{fig:freejoin}
An example of the transformation of a signature as the free end is
joined to a loop-end. In this case the source signature $(3,3,\{010011\})$
gives rise to the target $(2,5,\{0101\})$.}
\end{figure}

Open meanders are a little more complicated. The first part of the
necessary generalisation consists in adding an extra piece of information 
to our signature. We have to keep track of a single free end by specifying 
its position within the configuration of connected loop-ends. One simple
way of doing this is, in analogy with the infinite line, to specify the number,
$h_f-1$, of (connected) loop-ends below the free end, so that $h_f$ is
the position of the free end as counted from the bottom. So a configuration
is now described by a signature $(h,h_f,S)$. Naturally, we also have
to generalise the algorithm described above. We now start with the
signature $(1,1,0)$, that is a single free end below the infinite line,
and no crossings. The updating rules for adding a new crossing
are very similar to the ones described above for closed meanders.
One difference is that when a new loop is added (two loop-ends 
joined) below the free end, $h_f$ is increased (decreased) by 2. We also
need to consider what happens when joining the free end to a connected
loop-end. In this case we have to change the matching end of the
connected loop to the new free end in the target signature and we have to
change $h_f$ accordingly. An example illustrating this is shown in 
figure~\ref{fig:freejoin}. The updating rule when the free end does
not join with the loop-end on the other side is obviously just to increase 
(decrease) $h$ by 1 as the free end is moved below (above) the line.

\subsubsection{Semi-meanders}

Finally, in order to enumerate semi-meanders all we need do is
change the starting configuration. We now start in a position 
just before the first crossing of the semi-infinite line with $w$ loops
nested with one another. By running the algorithm for each
$w$ from 0 to $n$ we can count all semi-meanders with up to $n$
crossings. The generalisation to multi-component systems of semi-meanders 
is the same as for closed meanders.

\subsection{Computational complexity \label{sec:complex}}

\begin{figure}[h]
\begin{center}
\includegraphics[scale=0.7]{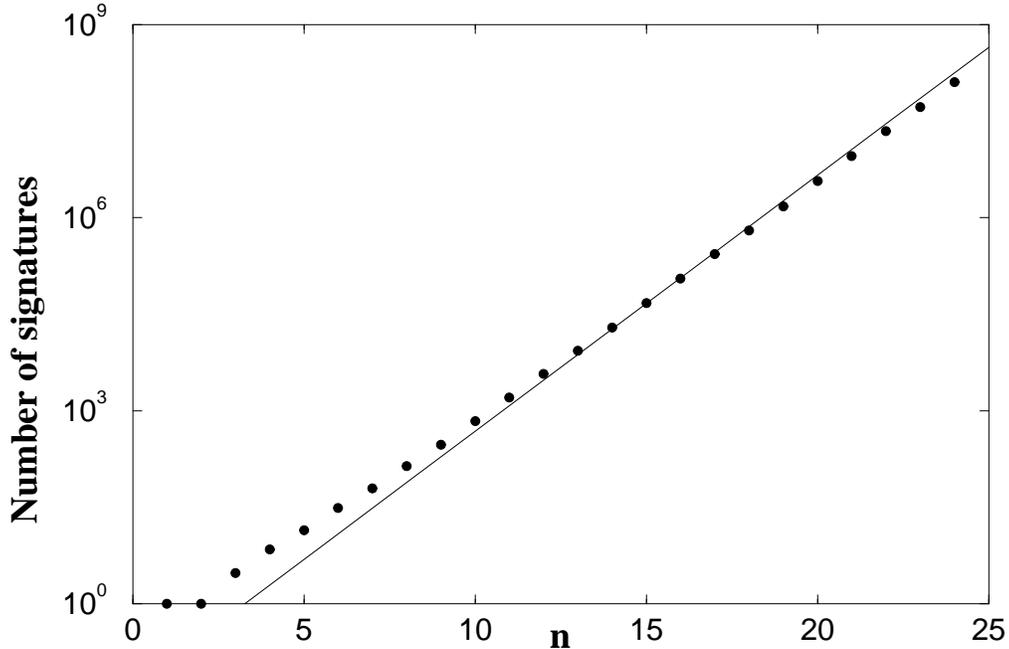}
\end{center}
\caption{\label{fig:numconf}
The number of distinct signatures required during the transfer matrix
calculation of the number of closed meanders with $2n$ crossings.
The solid line, drawn as a guide to the eye, corresponds to
a growth rate $\lambda = 2.5$.}
\end{figure}

Using the new algorithm we have calculated $M_n$ up to $n=24$ as compared 
to the previous best of $n=16$ obtained by V. R. Pratt \cite{Pratt}. To 
fully appreciate the advance it should be noted that the computational 
complexity grows exponentially, that is the time required 
to obtain $n$ term grows asymptotically as $\lambda ^n$. For direct
enumerations time is simply proportional to $M_n$ and thus
$\lambda = \lim_{n \rightarrow \infty} M_{n+1}/M_n \approx 12.26 $.
Thus extending the count of the meandric number from 16 to 24 by
direct counting would have required approximately 
$12.28^8 \simeq 5\times10^8$ as much CPU time as the calculation
of the first 16 terms.
The transfer matrix method employed in this paper is far more 
efficient. In figure~\ref{fig:numconf} we have plotted the maximum number 
of signatures required in order to calculate the number of closed meanders 
up to order $n$.  As can be seen, the number of signatures grows
exponentially with $n$, and the numerical evidence suggests that the 
computational complexity is such that $\lambda \approx 2.5$, which
obviously is a very significant improvement on direct counting.
The drawback of the transfer matrix method is that, since we
need to store all the different signatures, the memory requirement
of the algorithm also grows exponentially with growth rate $\lambda$,
whereas direct counting algorithms typically have memory requirements
which are linear in $n$. In fact it is exactly the memory requirement
which is the major limitation of the transfer matrix method. 
The calculations reported in this paper used up to 2Gb of memory and
typically took a few days of CPU time.

Di Francesco {et al.} \cite{FGJ2000} carried out a more detailed
analysis of the complexity of the algorithm as it applies to
multi-component systems of closed meanders and obtained the
estimate $N_{{\rm sig}} \sim a^n = (2.582603\ldots)^n$,
for the number of signatures, very close to the purely empirical
estimate obtained here.

Another way of gauging the improved efficiency is to note that
the calculations for semi-meanders carried out in \cite{FGG2} 
were ``done on the parallel Cray-T3D (128 processors) of the
CEA-Grenoble, with approximately 7500 hours $\times$ processors.''
Or in total about 100 years of CPU time. The equivalent calculations
can be done with the transfer matrix algorithm in about 15 minutes
on a single processor workstation!

\subsection{Further particulars}

Finally a few remarks of a more technical nature. The number of contributing 
configurations becomes very sparse in the total set of possible states along 
the boundary line and as is standard in such cases one uses a hash-addressing 
scheme \cite{Mehlhorn}. Since the integer coefficients occurring in the  
expansion become very large, the calculation was performed using modular 
arithmetic \cite{Knuth}. This involves performing the calculation modulo 
various prime numbers $p_i$ and then reconstructing the full integer
coefficients at the end. 
In calculations involving only single component meanders most of
the memory is used storing the signatures so we used primes
of the form $p_i=2^{30}-r_i$. The Chinese remainder theorem
ensures that any integer has a unique representation in terms of residues. If 
the largest integer occurring in the final expansion is $M$, then we 
have to use a number of primes $m$ such that $p_1p_2\cdots p_m > M$.  Up to 
3 primes were needed to represent the coefficients correctly.
In calculations involving systems of multi-component meanders the
main storage requirement comes from the generating functions. 
In order to save memory we used primes of the form 
$p_i=2^{15}-r_i$ so that the residues of the coefficients in the
polynomials could be stored using 16 bit integers. In this case we used
up to 6 primes.

\section{Conclusion\label{sec:conclusion}}

We have presented an improved algorithm for the enumeration of plane
meanders. The computational complexity of the algorithm for the
problem of closed meanders is estimated to 
be $2.5^n$, much better than direct counting algorithms which have 
complexity $12.26^n$. Implementing this algorithm enabled us to obtain 
closed meanders up to order $n=24$, compared to the previous best
of $n=16$. We also counted the number of open meanders up to order
$n=43$ and semi-meanders up to order $n=45$. From our extended series we 
obtained precise estimates for the connective constants and critical 
exponents \cite{JG2000}.
This showed that a recent conjecture for the exact value of the semi-meander 
critical exponent is unlikely to be correct, while the conjectured exponent 
value for closed and open meanders is just consistent with the results from 
the analysis.

\section*{E-mail or WWW retrieval of series}

The series for the various generating functions so far generated
with this algorithm and studied in \cite{JG2000}
can be obtained via e-mail by sending a request to 
I.Jensen@ms.unimelb.edu.au or via the world wide web on the URL
http://www.ms.unimelb.edu.au/\~{ }iwan/ by following the instructions.

\section*{Acknowledgements}

Financial support from the Australian Research Council is gratefully 
acknowledged.


\begin{thebibliography}{10}

\bibitem{LZ} S. K. Lando and A. K. Zvonkin, 
Theoret. Comput. Science {\bf 117}, 227 (1993).

\bibitem{Poincare} H. Poincar\'e, Rend. Circ. Mat. Palermo {\bf 33},
375 (1912).

\bibitem{Touchard} J. Touchard, Canad. J. Math. {\bf 2}, 385 (1950).

\bibitem{Koehler} J. E.  Koehler, J. Combin. Theory {\bf 5}, 135 (1968).

\bibitem{Lunnon} W. Lunnon, Math. Comp. {\bf 22}, 193 (1968).

\bibitem{Arnold} V. Arnold, Siberian Math. J., {\bf 29} 717 (1988).

\bibitem{KS} K. H. Ko and L. Smolinsky, 
Pacific J. Math. {\bf 149}, 319 (1991).

\bibitem{HMRT} K. Hoffmann, K. Mehlhorn, P. Rosenstiehl, and R. E. Tarjan,
Information and Control {\bf 68}, 170 (1988).

\bibitem{AM} N. Alon and W. Maass, 
J. Comput. System Sci. {\bf 37}, 118 (1988).

\bibitem{FGG1} P. Di Francesco, O. Golinelli and E. Guitter,
Math. Comput. Modelling {\bf 26}, 97 (1997).

\bibitem{FGG2} P. Di Francesco, O. Golinelli and E. Guitter,
Nucl. Phys. B {\bf 482}, 497 (1996).

\bibitem{FGG3} P. Di Francesco, O. Golinelli and E. Guitter,
Commun. Math. Phys. {\bf 186}, 1 (1997).

\bibitem{Francesco1}  P. Di Francesco, 
Commun. Math. Phys. {\bf 191}, 543 (1998);
J. Math. Phys. {\bf 38}, 5905 (1997).

\bibitem{Makeenko} Y. Makeenko, Nucl. Phys. Proc. Suppl. {\bf 49}, 226 (1996).

\bibitem{SS} G. W. Semenoff and R. J. Szabo, 
Int. J. Mod. Phys. {\bf A12}, 2135 (1997).

\bibitem{Francesco2}  P. Di Francesco, 
preprint http:://arxiv.org/abs/math-ph/9911002

\bibitem{Kholodenko} A. L. Kholodenko, 
J. Geom. Phys. {\bf 33}, 23 (2000).

\bibitem{Pratt} V. R. Pratt, in N. Sloane, 
{\em Sloane's On-Line Encyclopedia of Integer Sequences}, 
http://www.research.att.com/\~{}njas/sequences/

\bibitem{FGG4} P. Di Francesco, O. Golinelli and E. Guitter,
Nucl. Phys. B {\bf 570}, 699 (2000).

\bibitem{FGJ2000} P. Di Francesco,  E. Guitter and J. L. Jacobsen,
Nucl. Phys. B {\bf 580}, 757 (2000).

\bibitem{JG2000} I. Jensen and A. J. Guttmann, 
J. Phys. A {\bf 33}, L187 (2000).

\bibitem{Guttmann89} A. J. Guttmann, in {\em Phase Transitions and
Critical Phenomena}, Vol. 13, eds. C. Domb and J. L. Lebowitz,
Academic Press, New York (1989), pp 1-234.

\bibitem{Enting} I. G. Enting, J. Phys. A {\bf 13}, 3713 (1980).

\bibitem{Derrida} B. Derrida, J. Phys. A {\bf 14}, L5 (1981).

\bibitem{DV} M.-P. Delest and G. Viennot, 
Theoret. Comput. Sci. {\bf 34}, 169 (1984).

\bibitem{Mehlhorn} K. Mehlhorn, {\em Data Structures and Algorithms I:
Sorting and Searching}, EATCS Monographs on Theoretical Computer Science,
Springer-Verlag, Berlin (1984).

\bibitem{Knuth} D. E. Knuth, {\em Seminumerical Algorithms (The Art of
Computer Programming 2)}, Addison-Wesley, Reading, MA (1969). 


\end{thebibliography}
\end{document}